\documentclass[aps,prd,nofootinbib,onecolumn,floatfix,superscriptaddress]{revtex4-1}

\usepackage{natbib}			   
\usepackage{amsmath}                
\usepackage{amssymb}                
\usepackage{graphicx}
\usepackage{bm}

\usepackage[utf8]{inputenc} 

\usepackage{psfrag}
\usepackage{xspace}

\usepackage{ifpdf}
\ifpdf
\DeclareGraphicsExtensions{.pdf, .jpg, .tif}
\usepackage[
      pdftex,                  
      colorlinks=true,          
      urlcolor=blue,           
      menucolor=black,          
      linkcolor=blue, 
      anchorcolor=blue,  
      citecolor=blue,
      filecolor=blue,
      bookmarks=true,           
      bookmarksopen=true,       
      hyperfootnotes=false,     
      pdfpagemode=UseOutlines,  
      pdfauthor={The HERMES Collaboration},
      pdftitle={Kinematic Dependence of 
	Longitudinal Double-Spin Asymmetries at HERMES},
]{hyperref}
\else
\DeclareGraphicsExtensions{.eps, .jpg}
\usepackage[%
  breaklinks=true,%
  colorlinks=true,%
  linkcolor=blue,anchorcolor=blue,%
  citecolor=blue,filecolor=blue,%
  menucolor=blue,pagecolor=blue,%
  urlcolor=blue]{hyperref}
\fi

  \newcommand{\approptoinn}[2]{\mathrel{\vcenter{
  \offinterlineskip\halign{\hfil$##$\cr
    #1\propto\cr\noalign{\kern2pt}#1\sim\cr\noalign{\kern-2pt}}}}}

\newcommand{\xbj}{\ensuremath{x_{\mathrm{B}}}\xspace}

\newcommand{\de}{{\rm\,d}}

\newcommand{\ubar}{\ensuremath{\bar{u}}\xspace}
\newcommand{\dbar}{\ensuremath{\bar{d}}\xspace}
\newcommand{\sbar}{\ensuremath{\bar{s}}\xspace}

\begin{document}

\title{Reply to Comment on \\
 ``Reevaluation of the parton distribution of strange quarks in the nucleon'' }

\def\groupargonne{\affiliation{Physics Division, Argonne National Laboratory, Argonne, Illinois 60439-4843, USA}}
\def\groupbilbao{\affiliation{Department of Theoretical Physics, University of the Basque Country UPV/EHU, 48080 Bilbao, Spain}} \def\groupbilbaoi{\affiliation{IKERBASQUE, Basque Foundation for Science, 48013 Bilbao, Spain}}
\def\groupbnl{\affiliation{Brookhaven National Laboratory, Upton, New York 11772-5000, USA}}
\def\grouperlangen{\affiliation{Physikalisches Institut, Universit\"at Erlangen-N\"urnberg, 91058 Erlangen, Germany}}
\def\groupillinois{\affiliation{Department of Physics, University of Illinois, Urbana, Illinois 61801-3080, USA}}
\def\groupnone{\noaffiliation}

\groupbnl
\groupargonne
\groupillinois
\grouperlangen
\groupbilbao
\groupbilbaoi

\author{E.C.~Aschenauer}  \groupbnl
\author{H.E.~Jackson}  \groupargonne
\author{S.~Joosten}   \groupillinois
\author{K.~Rith}  \grouperlangen
\author{G.~Schnell}  \groupbilbao\groupbilbaoi 
\author{C.~Van~Hulse}  \groupbilbao 

\collaboration{On behalf of the HERMES Collaboration} \noaffiliation

\begin{abstract}
A Comment on the recently published reevaluation
of the polarization-averaged parton distribution of
strange quarks in the nucleon using final data on 
the multiplicities of charged kaons in semi-inclusive
deep-inelastic scattering (DIS) is reviewed. Important
features of the comparison of one-dimensional 
projections of the multidimensional HERMES data
are pointed out. A test of the leading-order
extraction of $\xbj  S(\xbj)$ using the difference between
charged-kaon multiplicities is repeated. The results
are consistent with leading-order predictions within the 
uncertainties in the input data, and do not invalidate the 
earlier extraction of $\xbj S(\xbj)$. 
\end{abstract}

\date{\today \ -- version 2.0}
\maketitle

\section{Introduction}

In the {\it Comment on ``Reevaluation of the parton distribution of strange quarks in the nucleon'' }, the Author presents a number of studies 
to conjecture that the analysis presented in Ref.~\cite{Airapetian:2013zaw} 
likely suffers from effects that invalidate the 
leading-order analysis used 
in that publication. 
For the studies presented part of the charged-hadron
multiplicity database of Ref.~\cite{Airapetian:2012ki} was used.

In our opinion, the Author
has drawn erroneous conclusions, using HERMES data on pion multiplicities~\cite{Airapetian:2012ki}, regarding the extraction of the 
strange-quark parton distribution $\xbj S(\xbj)$ from
charged-kaon multiplicities. 
In addition, his check of this extraction using the properties of the charged-kaon difference multiplicity and its derivative, the $K'$ ``multiplicity'' [cf.~Eq.~\eqref{eq:7lhs}], is invalidated by the 
large uncertainties in the values used
for relevant fragmentation functions (FFs) and parton distributions.
We present below the results from a repetition of
that analysis using a range of parton distribution sets.
The spread in those results precludes credible
conclusions and demonstrates the sensitivity of the analysis to
poorly known input data (unfavored FFs, strange-quark
 distributions, and mixed singlet and nonsinglet quantities).

Before going into detail we would like to recall some of the motivations
for the analysis presented in Ref.~\cite{Airapetian:2013zaw}:
\begin{itemize}
\item the strangeness distribution is a poorly known quantity:
  due to its usual suppression in most observables it is difficult to
  access;
\item charge- and flavor-separated FFs are, at this
  point, still not well constrained. Indeed, the FFs
  for kaons differ significantly in the various phenomenological
  extractions, employing strong assumptions to reduce the number
  of independent functions. One of the reasons for this is the
  dominance of $e^{+}e^{-}$ annihilation data in those extractions that does not easily
  allow for a charge separation of the FFs. For the analysis of the strange-quark distribution in Ref.~\cite{Airapetian:2013zaw} an isoscalar extraction was
  employed to avoid as much as possible this complication of involving badly constrained FFs, e.g.,
  only charge-summed FFs are needed from the global analysis
  of kaon FFs, an aspect that  
  validates the use of the FF phenomenology, which is dominated by  $e^{+}e^{-}$
  annihilation data; 
 \item we reiterate here again the statement in Ref.~\cite{Airapetian:2013zaw} that ``while a full next-to-leading order (NLO)  extraction would be preferred, such a procedure using semi-inclusive DIS data is not currently available,'' and that as such ``a leading-order extraction is an important first step''.
\end{itemize}

\section{Data analyses and comparisons}

The Author of the Comment~\cite{Stolarski:2014jka} offers a number of checks that in his opinion
should/could support or disprove the conclusions drawn in Ref.~\cite{Airapetian:2013zaw}.
In Sec.~III B the charged-pion sum is analyzed, which has little to do with 
the analysis presented in \cite{Airapetian:2013zaw}. 
As the Author points out himself,
because of $u$-quark dominance in pion fragmentation, the component of
the charged-pion multiplicity generated by strange-quark fragmentation should be small, e.g.,
only a few $\%$. Consequently, given currently attainable 
measurement precision, charged-pion multiplicities provide no information
on properties of strange quarks. It follows that the body of discussion in
Sec.~III concerning features of the leading-order (LO) description of the
pion multiplicity sum is not directly relevant to the extraction of $\xbj S(\xbj)$ from kaon 
data reported by HERMES.

\begin{figure}[*t]
	\centering
		\includegraphics[width=10cm]{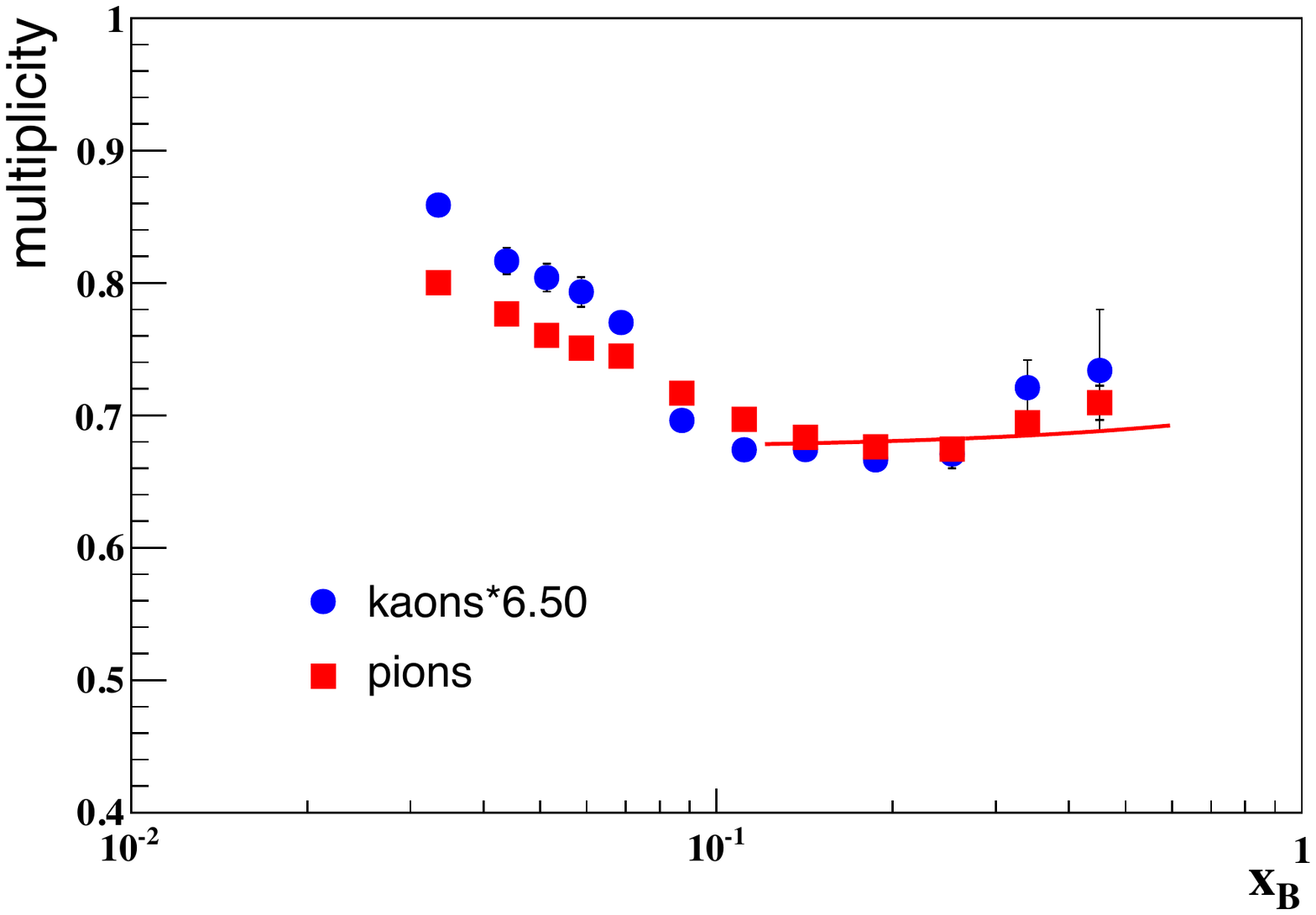}
	\caption{Comparison of the shapes of multiplicities corrected to 4$\pi$ 
         of charged kaons and pions
         in semi-inclusive DIS from a deuterium target, as a function
         of Bjorken \xbj. The kaon multiplicities are scaled to agree with those of pions in the range of \xbj where both distributions flatten.
         Data are plotted at the average \xbj of each individual \xbj bin.}
	\label{fig:mult_x}
\end{figure}

One salient point is the observation of ``the almost identical
shapes of the pion and kaon distributions.'' Our comparison of
the shapes is presented in Fig.~\ref{fig:mult_x} where the kaon values have
been renormalized to the pion values in the region of
$0.1<\xbj<0.45$ where both curves flatten, presumably due to
the absence of a contribution from strange-quark fragmentation.
While similar, the shapes are not nearly identical. 
While the similarity is an interesting observation, 
it may be coincidental, and likely can only be disentangled in a full QCD analysis, which is at this point still out of reach.
Until then possible implications on the LO extraction of $\xbj S(\xbj)$ presented in Ref.~\cite{Airapetian:2013zaw} cannot be assessed.

\begin{figure}[*t]
	\centering
		\includegraphics[width=10cm]{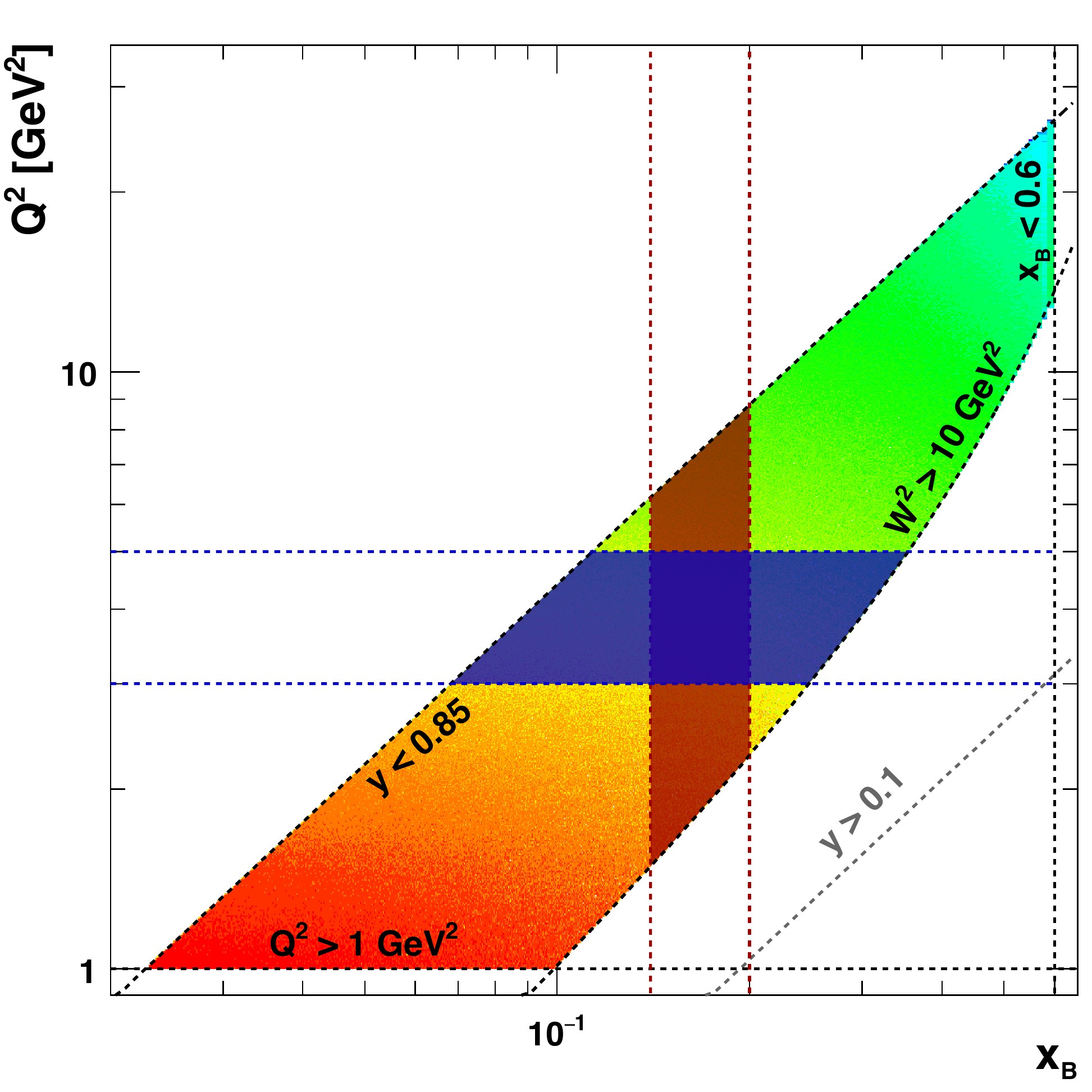}
	\caption{Born space in (\xbj,Q$^2$) corresponding to the multiplicities reported in Ref.~\cite{Airapetian:2012ki}. Kinematic regions covered by two data points with similar average kinematics, as discussed in the Comment~\cite{Stolarski:2014jka}, are superimposed (highlighted slices in either \xbj or Q$^2$).  
	Note that the color coding is logarithmic, 
	and that most of the events [\(\mathcal{O}(70\%)\)] in either of the two bins are in fact not shared.
	\label{fig:xQ2plane}
	}
\end{figure}

For all the discussion, a grave conceptual misunderstanding must be pointed out,
also because it appears to cloud the analyses~\cite{Leader:2015hna} referred to by the 
Author to strengthen his case. Although not related to the kaon multiplicities 
analyzed in Ref.~\cite{Airapetian:2013zaw}, 
the Author indulges in {\em invalid comparisons} of pion multiplicity values 
(last paragraph of Sec.~III A).
Different one-dimensional projections of a multiparameter observable,
such as a multiplicity, are not topologically equivalent. In the case
discussed in the Comment~\cite{Stolarski:2014jka}, the corresponding sections in the \xbj and Q$^2$ projections
of the observable span different, albeit, overlapping regions of the
Born space accepted in the measurement. From setting the values of the
observables drawn from these two different sections equal 
[e.g., erroneously assuming that the average multiplicity should correspond to the multiplicity at (event-weighted) average kinematics],
the unjustified conclusion is drawn that the measurements presented in 
Ref.~\cite{Airapetian:2012ki} are not consistent.  
In contrast to the Author's claim that ``a large part of the data [in the two representations] are the same'', Fig.~\ref{fig:xQ2plane} clearly demonstrates the contrary: only a small part
of the kinematic regions covered by those two data points do overlap.

As an example, multiplicities in the $\xbj$-$z$ representation involve integrations over Q$^2$ (as well as $P_{h\perp}$). Consequently, when comparing to theoretical predictions, the same integration has to be performed, e.g., for the LO parton-model expression one has to evaluate
\begin{equation}
\mathcal{M}^{\pi} (\xbj,z)
 =\frac{{\sum}_q e_q^2 \int_{\text{Q}^{2}_{\text{min}}(\xbj)}^{\text{Q}^{2}_{\text{max}}(\xbj)}
q(\xbj ,\text{Q}^2) D_q^{\pi}(z,\text{Q}^2) \de \text{Q}^{2}}
{{\sum}_q e_q^2\int_{\text{Q}^{2}_{\text{min}}(\xbj)}^{\text{Q}^{2}_{\text{max}}(\xbj)} q(\xbj ,\text{Q}^2)\de \text{Q}^{2}},
\label{eq:leading}
\end{equation}
or an analogous integration for any other type.

The Author points out that the (very) low-\xbj region can be neglected in the discussion of the 
two bins in question ($3~\text{GeV}^{2} < Q^{2}< 5~\text{GeV}^{2}$ of the $z$--$Q^{2}$ representation, and $0.14<\xbj<0.2$ of the $z$--$\xbj$ representation in Ref.~\cite{Airapetian:2012ki}), as the relevant Q$^2$ bin is large enough not to include the low-\xbj data. Indeed, \xbj is larger than about 0.07. But already from Fig.~1 of the Comment~\cite{Stolarski:2014jka} it should have been clear that the average \xbj in that bin sits basically at the minimum of the multiplicity distribution, such that the multiplicity averaged over  that $\xbj$ range ($0.07 \lesssim \xbj \lesssim 0.35$) must be larger than the multiplicity at that average \xbj, as it is indeed found.

This clarification of how to use the multiplicity database~\cite{Airapetian:2012ki} is an important enough aspect for {\em any} analysis of these multiplicities, especially when attempting to compare to theory predictions, such that it deserves to be again highlighted here.

In Sec.~IV, the Author turns from the isoscalar extraction to an
analysis involving the difference between $K^{+}$ and $K^{-}$ multiplicities, e.g.,
\begin{align}
\frac{\de N^{K^{\text{diff}}}}{\de N^{\text{DIS}}} ~~~ & \equiv ~~~ \frac{\de  ( N^{K^{+}} - N^{K^{-}} ) }{\de N^{\text{DIS}}} \label{eq:5lhs}\\
&\!\!\!\! \stackrel{\text{LO}, s=\sbar}{=}   \frac{ (u_{v}+d_{v}) ( 4 D_{u}^{K^{+}} - 4 D_{\ubar}^{K^{+}} + D_{d}^{K^{+}} - D_{\dbar}^{K^{+}} ) } {5Q + 2S}, \label{eq:5rhs}
\end{align}
where the second line is obtained at LO assuming $s(\xbj) = \sbar(\xbj)$, and where $Q\equiv u+\ubar + d+\dbar$ and $S\equiv s+\sbar$ are as defined in Ref.~\cite{Airapetian:2013zaw}, $D_{q}^{K^{+}}$ are the quark-flavor separated FFs\footnote{It is implicitly assumed in this text that FFs are integrated over the relevant range in the fractional hadron energy $0.2<z<0.8$.} into $K^{+}$, and $u_{v}$ ($d_{v}$) the up (down) valence distributions.

First of all, it should be noted that the difference multiplicity is inherently
nonsinglet and consequently sensitive to the flavor structure of 
the fragmentation process. Also, the features of the Q$^2$ evolution
of this quantity is different than that of the sum multiplicity.
This applies even more to the so-called $K'$ ``multiplicity'', a mixture of the nonsinglet difference and the isoscalar charge-sum multiplicity ($\de  N^{K }/\de N^{\text{DIS}}$):
\begin{align}
\frac{\de N^{K'}}{\de N^{\text{DIS}}} ~~ & \equiv  ~~ \frac{5Q+2S}{Q} \frac{\de   N^{K} }{\de N^{\text{DIS}}} - \frac{5Q+2S}{ u_{v}+d_{v}} \frac{\de   N^{K^{\text{diff}}} }{\de N^{\text{DIS}}} \label{eq:7lhs}
 \\
&\!\!\!\!\! \stackrel{\text{LO}, s=\sbar}{=}   8 D_{\ubar}^{K^{+}} + 2 D_{\dbar}^{K^{+}}   + \frac{S}{Q} D_{S}^{K}  \label{eq:7rhs}
\\
& = ~~\, 8 D_{{u}}^{K^{-}} + 2 D_{{d}}^{K^{-}}   + \frac{S}{Q} D_{S}^{K}
\, .
\end{align}

In the Comment~\cite{Stolarski:2014jka} an extraction of the kaon charge-difference multiplicity 
(and of the  $K'$ ``multiplicity'')
is carried out using MSTW08~\cite{Martin:2009iq} and alternatively NNPDF3.0~\cite{Ball:2014uwa} LO parton-distribution functions (PDFs) 
as well as a determination of the relevant nonstrange
fragmentation functions ``from the data at high \xbj, exactly as was done
by HERMES for $D_Q^K$''. The Author states that contrary to expectation
the observable decreases at low \xbj, and that ``it is hard
to expect that the Q$^2$ dependence of $D_{\ubar}^{K^+}+D_{\dbar}^{K^+}$ can
fully explain the shape.'' This is concluded to be ``an indication of the failure
of the conventional LO pQCD parton model approach.'' 

\begin{figure}[*t]
	\centering
    \includegraphics[width=0.5\textwidth]{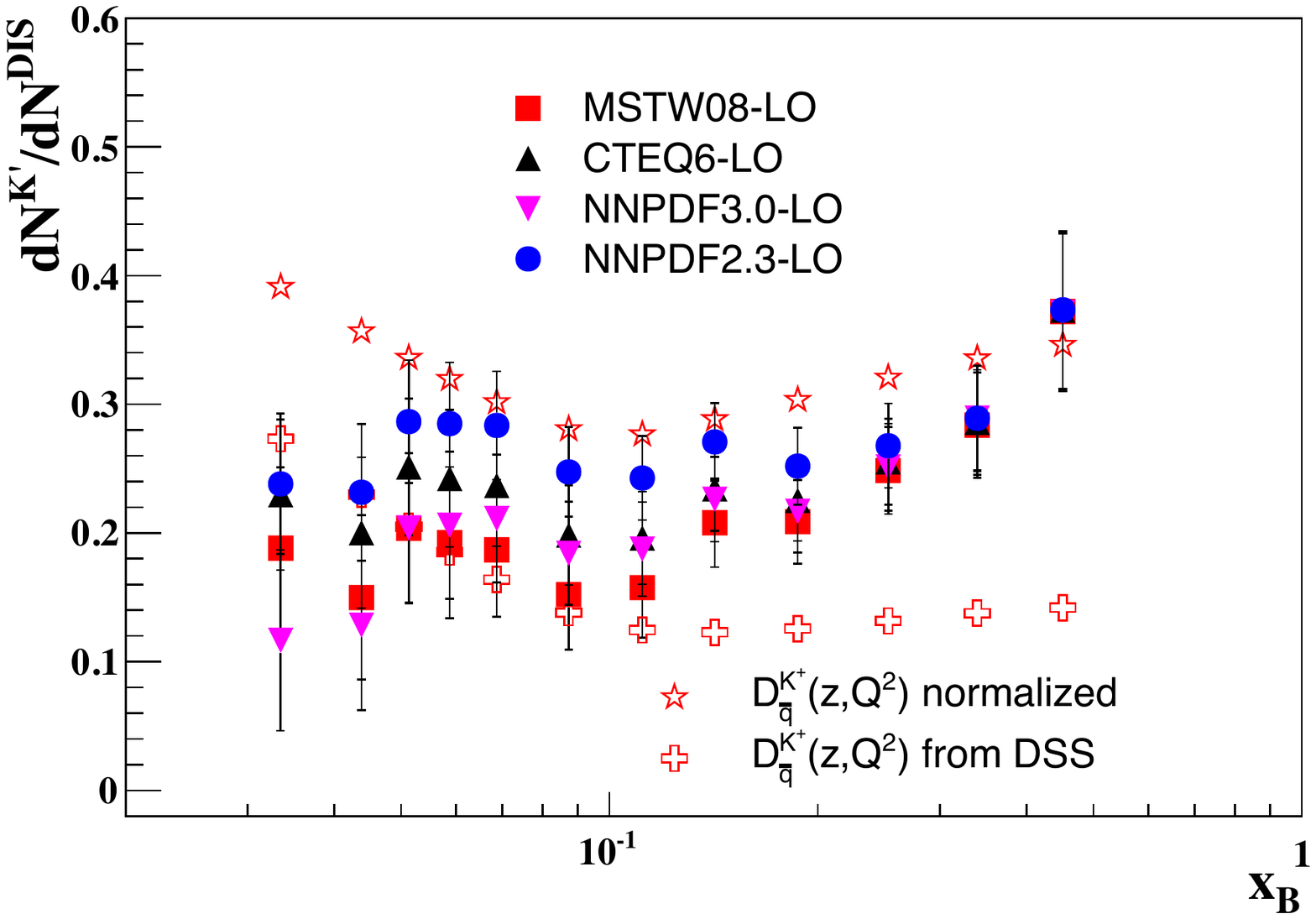}~\includegraphics[width=0.5\textwidth]{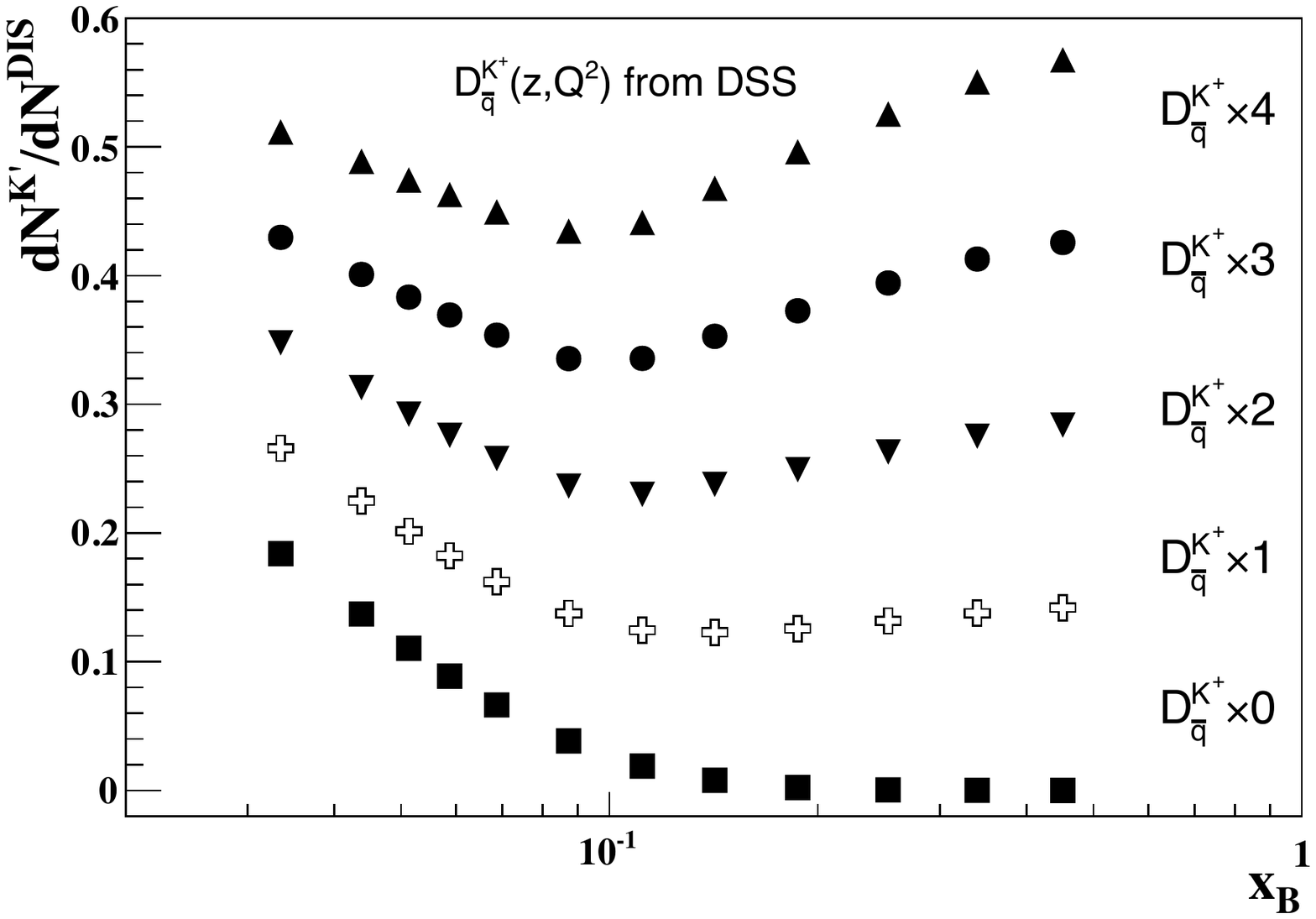}
	\caption{Left: $K'$ ``multiplicities'' corrected to 4$\pi$ 
         of charged kaons  
         in semi-inclusive DIS from a deuterium target, as a function
         of \xbj, evaluated with Eq.~\eqref{eq:7lhs} using MSTW08 (squares), CTEQ6 (upward triangles), NNPDF3.0 (downward triangles), and NNPDF2.3 (full circles) LO PDF sets. 
Also shown are the LO predictions [e.g., Eq.~\eqref{eq:7rhs}] using DSS FFs (crosses) or using the high-\xbj HERMES data to constrain the unfavored $\ubar$ and $\dbar$ to $K^{+}$ FFs (stars). For both the crosses and stars, $S\!\int\! \de z D_{S}^{K}$ from Ref.~\cite{Airapetian:2013zaw} and $Q$ from NNPDF3.0 were used. Note that uncertainties on PDFs or FFs---when available at all---were not propagated, but only total experimental uncertainties. 
Right:
Equation~\eqref{eq:7rhs} evaluated in the same way as the crosses to the left, but for a range of {\em scaled} DSS unfavored nonstrange kaon FFs [using scaling factors from 0 (bottom) to 4 (top) as in Fig.~2 (left) in the Comment]. As in the Comment~\cite{Stolarski:2014jka}, all points were evaluated at the average Q$^{2}$ of each individual \xbj bin.
         }
	\label{fig:mult_dif}
\end{figure}

We have repeated such analysis using a set of LO PDFs, namely NNPDF3.0 and MSTW08 (as in the Comment~\cite{Stolarski:2014jka}) as well as CTEQ6~\cite{Pumplin:2002vw} and NNPDF2.3~\cite{rojo:nnpdf2.3lo}. These were then used to evaluate the $K'$ multiplicities according to Eq.~\eqref{eq:7lhs}.
In addition, we have followed the Author's 
suggestion of using the ``high-\xbj limit'' of $\de N^{K'}/\de N^{\text{DIS}}$ to normalize 
the unfavored fragmentation combination $(8D_{\ubar}^{K^+}+2D_{\dbar}^{K^+})$,
and used the data base of Ref.~\cite{deFlorian:2007aj} (DSS) to generate the Q$^2$ dependence of those.\footnote{Here we deviate from the path of the Author by not using {\sc QCDNUM} for performing the Q$^2$ evolution.} 
(The ``high-\xbj limit'' was obtained by evaluating in the last \xbj bin a polynomial fit to the NNPDF3.0 data points for $\xbj>0.1$.)
Alternatively, the original DSS FFs~\cite{deFlorian:2007aj} were used.\footnote{It should
be noted that the DSS analysis is based, in part, on a preliminary unpublished version of
the HERMES data on charged-meson multiplicities from a hydrogen target.}
Together with $S\!\int \!\de z D_{S}^{K}$ from Ref.~\cite{Airapetian:2013zaw} and $Q$ from NNPDF3.0 these  two sets of FFs were used to evaluate Eq.~\eqref{eq:7rhs}. 
For the Q$^2$ evolution of the PDFs the package LHAPDF~\cite{Buckley:2014ana} was used.
Our results are presented in Fig.~\ref{fig:mult_dif}.  

In contrast to the results in the Comment~\cite{Stolarski:2014jka}, one can find a reasonable
agreement between the HERMES extraction of $\de N^{K'}/\de N^{DIS}$ and the latter
leading-order prediction, when uncertainties in the FFs and PDFs are accounted for.

\begin{figure}[*t]
	\centering
    \includegraphics[width=9.8cm]{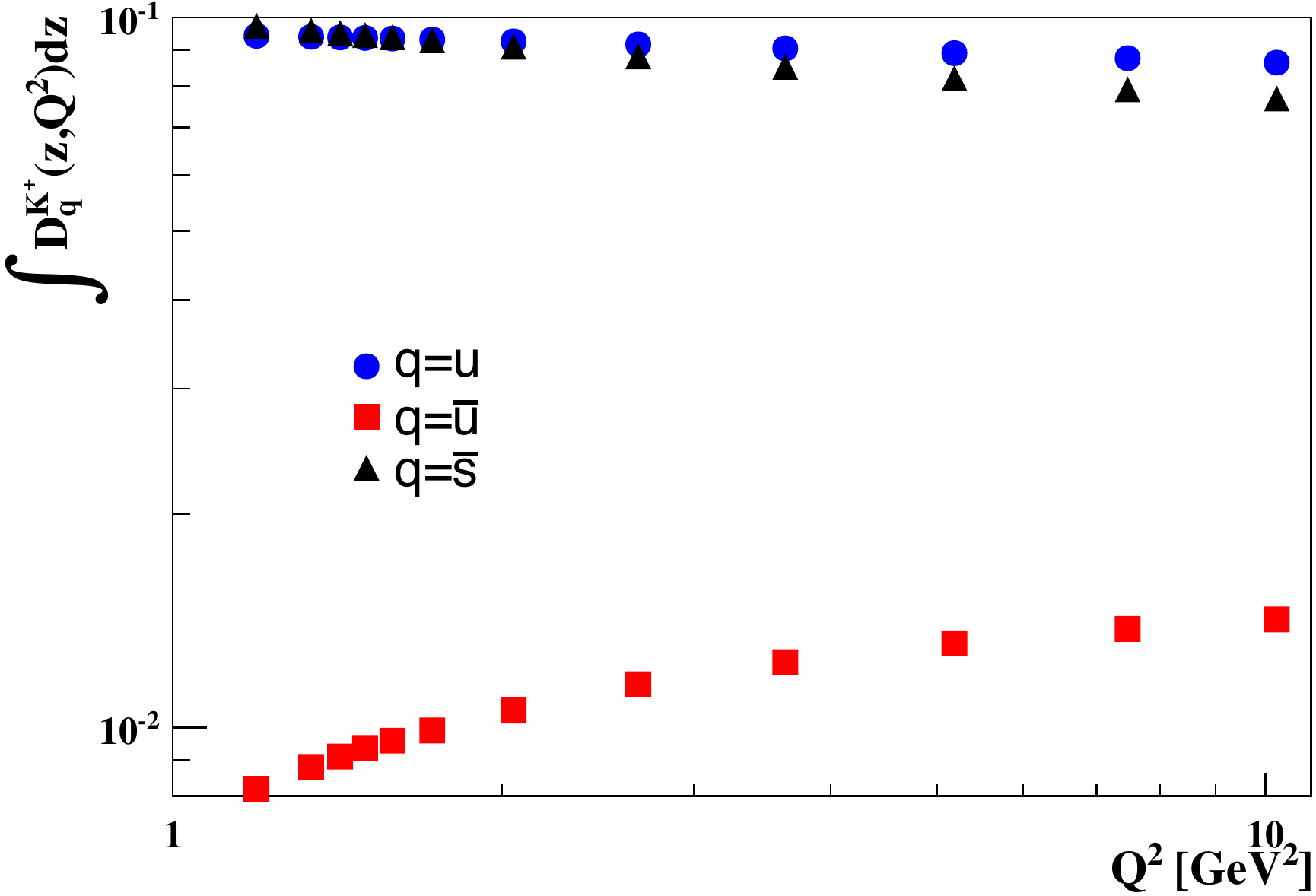}
	\caption{The Q$^2$ dependence of the DSS favored $u$ and $\sbar$ (scaled by a factor 1/7) as well as the unfavored $\ubar$ to $K^{+}$ fragmentation functions. Note that for the DSS FF set $D_{\ubar}^{K^{+}} = D_{\dbar}^{K^{+}} $.
	Chosen Q$^{2}$ values correspond to the average Q$^{2}$ of the \xbj bins in Figs.~\ref{fig:mult_dif} and \ref{fig:mult_scal}.
	}
	\label{fig:DSSq2}
\end{figure}

The following observations are worthwhile mentioning:
\begin{itemize}
\item first it should be noted that the DSS predictions for the kaon FFs, particularly for
the unfavored FFs are, to say the least, uncertain. In the DSS 
compilation all the kaon unfavored FFs are assumed to be equal, and in particular $D_{\ubar}^{K^{+}}=D_{\dbar}^{K^{+}}$. Taking into account uncertainties in the DSS compilation of the unfavored nonstrange kaon FFs~\cite{Epele:2012vg}, a broad range of ``LO predictions''  based on Eq.~\ref{eq:7rhs} is obtained filling essentially the range between the squares and circles in the right of Fig.~\ref{fig:mult_dif};
\item the two choices for the unfavored nonstrange $K^{+}$ FFs, namely DSS (crosses in Fig.~\ref{fig:mult_dif} left) and the ones derived from the ``high-\xbj limit'', e.g., the last \xbj bin (stars in Fig.~\ref{fig:mult_dif} left), envelop the region populated by the HERMES data folded with various LO PDF sets; 
\item as expected, the results for the extracted values of $\de N^{K'}/\de N^{\text{DIS}}$ converge
to a common locus for $\xbj>0.1$,
since for negligible sea-quark contributions (common among the chosen PDF sets) Eq.~\eqref{eq:7lhs} reduces simply to ten times the $K^{-}$ multiplicity\footnote{It is interesting to note that the DSS prediction in this region significantly undershoots the data pointing to underestimated strength in the unfavored nonstrange kaon FFs.}, independent of the values for the valence distributions.
In the region $\xbj<0.1$ the spread in the 
extracted values of $\de N^{K'}/\de N^{\text{DIS}}$ becomes large reflecting the large
variation in the sea-quark PDFs over the four sets used. While NNPDF2.3 results in a reasonable description when compared to the set of unfavored nonstrange FFs derived from the ``high-\xbj limit,'' the overall choices of PDF and FF sets result in a wide range of ``predictions'' for either side of Eq.~(7) of the Comment~\cite{Stolarski:2014jka}, i.e., Eqs.~\eqref{eq:7lhs} and \eqref{eq:7rhs} herein. This clearly demonstrates the large sensitivity of these nonsinglet quantities to subtleties in the choice of PDFs and FFs;
\item using the Q$^2$ evolution of the DSS FF set, we cannot reproduce the shapes of the curves in the Comment's Fig.~2 (left)~\cite{Stolarski:2014jka}. We present the quantity of Eq.~\eqref{eq:7rhs} on the right of Fig.~\ref{fig:mult_dif}. There is a strong underlying Q$^2$ dependence that leads to the rise of the $K'$ ``multiplicity'' with \xbj for large values of the unfavored FFs, something not visible in the version of this figure in the Comment~\cite{Stolarski:2014jka}. The strong Q$^2$ dependence of the unfavored DSS kaon FFs is depicted also in Fig.~\ref{fig:DSSq2};
\item even more striking than the lack of the high-\xbj rise for the considerably increased DSS FFs is the behavior of the lowest curve in the corresponding figure of the Comment~\cite{Stolarski:2014jka}: when the unfavored FFs are set to zero, the \xbj behavior is driven entirely by the last term in Eq.~\eqref{eq:7rhs}. $S D_{S}^{K}$ rapidly reaches zero above $\xbj=0.1$~\cite{Airapetian:2013zaw}. As $Q$ remains sizable in that range of \xbj also the ratio $\frac{S}{Q} D_{S}^{K} $ should approach zero (as it does on the right of Fig.~\ref{fig:mult_dif} for the corresponding squares). In Fig.~2 left of the Comment~\cite{Stolarski:2014jka}, however, instead of going to zero that curve rises with \xbj above \xbj = 0.1. We have no explanation for this apparent discrepancy; 
\item the rapidly rising values as \xbj increases make extracting
a ``high-\xbj limit,'' as recommended in the Comment~\cite{Stolarski:2014jka}, very tenuous at best. 
Unlike the situation with the original isoscalar extraction of $S(\xbj)$~\cite{Airapetian:2013zaw}, where a high-\xbj limit  was used to constrain a  favored combination of FFs and where the multiplicity levels out and thus has a meaningful limit,
it appears that the notion of a ``high-\xbj limit'' is not applicable in this case, where both the last term in Eq.~\eqref{eq:7rhs} and the unfavored FFs in that equation are suppressed quantities. 

\end{itemize}

The large spread in the PDF-dependent extractions of $\de N^{K'}/\de N^{\text{DIS}}$ as
well as the spread between the predictions for this quantity using the 
values of the unfavored FFs from DSS and those from 
a nominal ``high-\xbj limit'' argue against using this quantity as a 
benchmark for testing the validity of the LO formalism.

\begin{figure}[*t]
	\centering
    \includegraphics[width=9.8cm]{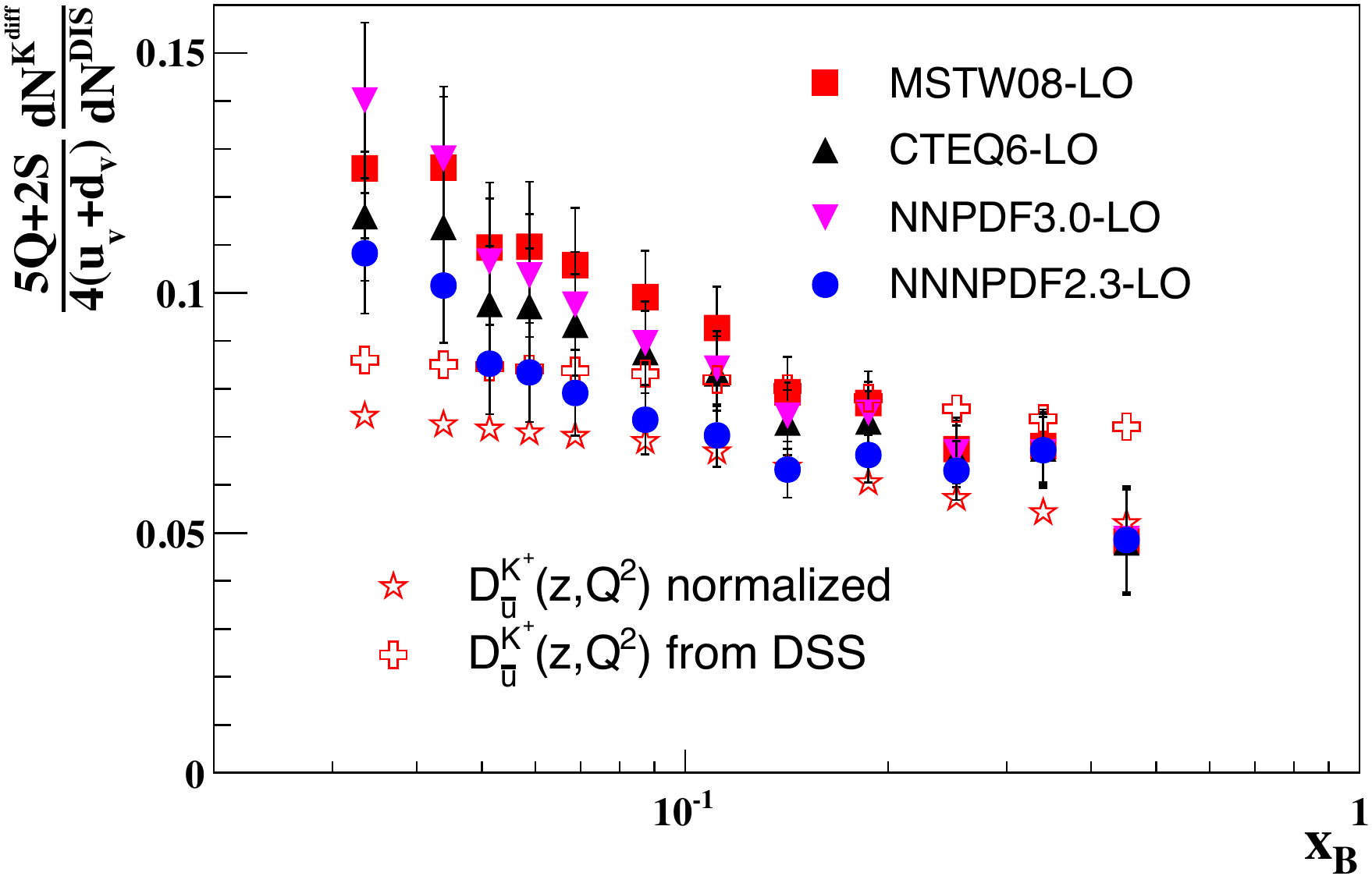}
	\caption{The scaled charge-difference multiplicities           
	$\de N^{K^{\text{diff}}}/\de N^{\mathrm{DIS}}\cdot (5Q+2S)/[4(u_v+d_v)]$  
	of charged kaons in semi-inclusive DIS from a deuterium target, as a function
         of \xbj, evaluated using various LO PDF sets [MSTW08 (squares), CTEQ6 (upward triangles), NNPDF3.0 (downward triangles), and NNPDF2.3 (circles)], 
          compared with the LO predictions based on solely DSS FFs (crosses) and DSS FFs combined with unfavored kaon FFs constrained using HERMES data at high \xbj (stars).
          As in the Comment~\cite{Stolarski:2014jka}, all points were evaluated at the average Q$^{2}$ of each individual \xbj bin.
          } 
	\label{fig:mult_scal}
\end{figure}

We have also extracted the quantity [cf.~Eq.~\eqref{eq:5rhs}]
\begin{align}
\frac{5Q+2S}{4(u_v+d_v)}    \frac{\de N^{K^{\text{diff}}}}{\de N^{\mathrm{DIS}}} 
& \stackrel{\text{LO}, s=\sbar}{=}   4 \left( D_{u}^{K^{+}} - D_{\ubar}^{K^{+}}\right)  + \left( D_{d}^{K^{+}} - D_{\dbar}^{K^{+}} \right) \label{eq:5scaled}
\end{align}
that is shown in the right panel of Fig.~2 of the Comment~\cite{Stolarski:2014jka} with
the same assumptions used in the extraction above. 
Four sets of LO PDFs were used again in the calculation of the quantity.
The result is presented in Fig.~\ref{fig:mult_scal} together with two LO predictions in terms of sums of FFs.
In contrast to the result shown in the Comment~\cite{Stolarski:2014jka} the agreement of the measurement
with the LO predictions is
reasonable except at the low \xbj values, 
and in view of the crudeness of the 
data we have for kaon FFs, perhaps as good as we can expect at this time.
Most striking is the spread originating from the use of the various PDF sets. 
This poses a serious concern about the sensitivity of this quantity to available PDF sets.
The corresponding plot in the Comment~\cite{Stolarski:2014jka} thus appears to be an  exaggeration of
the disagreement between LO prediction and data.

\section{Conclusion}

In contrast to the kaon multiplicities, those of the pions cannot be
satisfactorily described in LO. Until this problem is understood, it is difficult to relate
features of the pion data to the behavior of the kaon data.
Our studies involving charged-kaon difference multiplicities do not corroborate
the results presented in the Comment~\cite{Stolarski:2014jka}. Rather we find, within the uncertainties dictated
by the current limited knowledge of strange FFs and even LO PDF sets, that the difference data
is consistent with LO predictions.

We reiterate again the statement from Ref.~\cite{Airapetian:2013zaw} that
while a NLO extraction of $S(\xbj)$ would be preferred, such a procedure
using SIDIS data is not currently feasible, and a LO extraction is
an important first step.

\acknowledgments
This work was supported by the German Bundesministerium f\"ur Bildung und 
Forschung (BMBF), and the Deutsche Forschungsgemeinschaft (DFG); the 
Basque Foundation for Science (IKERBASQUE) and the UPV/EHU under program 
UFI 11/55; as well as the U.S. Department of Energy (DOE) and the National 
Science Foundation (NSF).

\section{Appendix: Tables of multiplicities used}
Tables~\ref{tab:pip}--\ref{tab:kp}, and \ref{tab:km} list the multiplicity values used herein including the average values of \xbj and Q$^{2}$ for $\pi^{+}$, $\pi^{-}$, $K^{+}$, and $K^{-}$, respectively. The total uncertainties on the multiplicities, obtained by adding in quadrature both statistical and systematic uncertainties, are propagated for all quantities computed, in particular, the multiplicity differences and sums.

\begin{table}[t]
\caption{Multiplicities corrected to 4$\pi$ of $\pi^{+}$ mesons in semi-inclusive DIS from a deuterium target, as a function of Bjorken \xbj.}
\label{tab:pip}
\begin{tabular}{ccccc}
\hline\hline
~~~~~~$\langle \xbj\rangle~~~~~~$ & ~~~~$\langle$Q$^{2} \rangle$ $[\text{GeV}^{2}]$~~~~ & ~~~~Multiplicity~~~~ & ~~~~Statistical uncertainty~~~~ & ~~~~Systematic uncertainty \\
\hline
0.033 	& 1.19 	& 0.422 & 0.002 & 0.010 \\
0.044 	& 1.34 	& 0.420 & 0.003 & 0.010 \\
0.051 	& 1.42 	& 0.408 & 0.003 & 0.008 \\
0.059 	& 1.50 	& 0.405 & 0.003 & 0.008 \\
0.069 	& 1.59 	& 0.406 & 0.002 & 0.007 \\
0.087 	& 1.73 	& 0.394 & 0.001 & 0.005 \\
0.112 	& 2.05 	& 0.382 & 0.002 & 0.006 \\
0.142 	& 2.67 	& 0.378 & 0.002 & 0.005 \\
0.187 	& 3.63 	& 0.378 & 0.002 & 0.004 \\
0.253 	& 5.19 	& 0.379 & 0.003 & 0.005 \\
0.340 	& 7.48 	& 0.389 & 0.005 & 0.005 \\
0.451 	& 10.23 	& 0.393 & 0.010 & 0.005 \\

\hline\hline
\end{tabular}
\end{table}

\begin{table}
\caption{Multiplicities corrected to 4$\pi$ of $\pi^{-}$ mesons in semi-inclusive DIS from a deuterium target, as a function of Bjorken \xbj.}
\label{tab:pim}
\begin{tabular}{ccccc}
\hline\hline
~~~~~~$\langle \xbj\rangle~~~~~~$ & ~~~~$\langle$Q$^{2} \rangle$ $[\text{GeV}^{2}]$~~~~ & ~~~~Multiplicity~~~~ & ~~~~Statistical uncertainty~~~~ & ~~~~Systematic uncertainty \\
\hline
0.033 	& 1.19 	& 0.378 & 0.002 & 0.008 \\
0.044 	& 1.34 	& 0.357 & 0.002 & 0.007 \\
0.051 	& 1.42 	& 0.353 & 0.003 & 0.006 \\
0.059 	& 1.50 	& 0.346 & 0.003 & 0.006 \\
0.069 	& 1.59 	& 0.339 & 0.002 & 0.006 \\
0.087 	& 1.73 	& 0.323 & 0.001 & 0.005 \\
0.112 	& 2.05 	& 0.315 & 0.002 & 0.004 \\
0.142 	& 2.67 	& 0.306 & 0.002 & 0.004 \\
0.187 	& 3.63 	& 0.298 & 0.002 & 0.003 \\
0.253 	& 5.19 	& 0.296 & 0.002 & 0.003 \\
0.340 	& 7.48 	& 0.305 & 0.004 & 0.004 \\
0.451 	& 10.23 	& 0.316 & 0.009 & 0.004 \\

\hline\hline
\end{tabular}
\end{table}

\begin{table}
\caption{Multiplicities corrected to 4$\pi$ of $K^{+}$ mesons in semi-inclusive DIS from a deuterium target, as a function of Bjorken \xbj.}
\label{tab:kp}
\begin{tabular}{ccccc}
\hline\hline
~~~~~~$\langle \xbj\rangle~~~~~~$ & ~~~~$\langle$Q$^{2} \rangle$ $[\text{GeV}^{2}]$~~~~ & ~~~~Multiplicity~~~~ & ~~~~Statistical uncertainty~~~~ & ~~~~Systematic uncertainty \\
\hline
0.033 	& 1.19 	& 0.084 	& 0.001 & 0.004 \\
0.044 	& 1.34 	& 0.083 	& 0.001 & 0.004 \\
0.051	& 1.42 	& 0.080 	& 0.001 & 0.004 \\
0.059	& 1.50 	& 0.080 	& 0.001 & 0.004 \\
0.069 	& 1.59 	& 0.079 	& 0.001 & 0.004 \\
0.087 	& 1.73 	& 0.073 	& 0.001 & 0.003 \\
0.112 	& 2.05 	& 0.073 	& 0.001 & 0.003 \\
0.142 	& 2.67 	& 0.072 	& 0.001 & 0.003 \\
0.187 	& 3.63 	& 0.074 	& 0.001 & 0.003 \\
0.253 	& 5.19 	& 0.074 	& 0.001 & 0.004 \\
0.340 	& 7.48 	& 0.080 	& 0.003 & 0.004 \\
0.451 	& 10.23 	& 0.075 	& 0.006 & 0.004 \\

\hline\hline
\end{tabular}
\end{table}

\begin{table}
\caption{Multiplicities corrected to 4$\pi$ of $K^{-}$ mesons in semi-inclusive DIS from a deuterium target, as a function of Bjorken \xbj.}
\label{tab:km}
\begin{tabular}{ccccc}
\hline\hline
~~~~~~$\langle \xbj\rangle~~~~~~$ & ~~~~$\langle$Q$^{2} \rangle$ $[\text{GeV}^{2}]$~~~~ & ~~~~Multiplicity~~~~ & ~~~~Statistical uncertainty~~~~ & ~~~~Systematic uncertainty \\
\hline
0.033 	& 1.19 	& 0.048 	& 0.001 	& 0.002 \\
0.044 	& 1.34 	& 0.043 	& 0.001 	& 0.002 \\
0.051 	& 1.42 	& 0.044 	& 0.001 	& 0.002 \\
0.059 	& 1.50 	& 0.042 	& 0.001 	& 0.002 \\
0.069 	& 1.59 	& 0.040 	& 0.001 	& 0.002 \\
0.087 	& 1.73 	& 0.034 	& 0.000 	& 0.002 \\
0.112 	& 2.05 	& 0.031 	& 0.001 	& 0.002 \\
0.142 	& 2.67 	& 0.032 	& 0.001 	& 0.002 \\
0.187 	& 3.63 	& 0.029 	& 0.001 	& 0.001 \\
0.253 	& 5.19 	& 0.029 	& 0.001 	& 0.002 \\
0.340 	& 7.48 	& 0.031 	& 0.002 	& 0.002 \\
0.452 	& 10.24 	& 0.038 	& 0.004 	& 0.002 \\

\hline\hline
\end{tabular}
\end{table}

\bibliographystyle{apsrev4-1}
\bibliography{multpaper}

\end{document}